# Monte Carlo simulation of $^{192}$Ir radioactive source in a phantom designed for brachytherapy dosimetry and source position evaluation

Samuel Chiquita[1]

[1]University of Porto, Porto, Portugal

## Abstract

In this report simulations of $^{192}$Ir source located inside a phantom designed for measuring the absorbed dose and radioactive source position are presented. Monte Carlo simulations were performed and results were compared with a theoretical model that enables to determine radioactive source position. A good fit from the simulated data to the theoretical model was obtained. Results show that Monte Carlo simulations allow to evaluate source position and absorbed dose with implications in treatment planning.

## Introduction

Brachytherapy treatments can be employed to destroy tumors while sparing healthy tissues. Iridium ($^{192}$Ir) radioactive source is routinely used in brachytherapy treatments. Several parameters can be used to characterize brachytherapy sources, namely radial dose function and anisotropy function (Nath et al., 1995). In order to ensure an accurate tumor treatment source position and dose determination are essential. Consequently, quality control shall be performed allowing to compare dose measurements with dose values calculated by the treatment planning system, to ensure that the prescribed dose to the target volume is within certain limits. Iridium-192 high dose rate (HDR) quality assurance of source position is performed with various methods, such as with radiochromic films and diamond detectors (Evans, Devic, & Podgorsak, 2007; Nakano et al., 2003; Nakano, Suchowerska, McKenzie, & Bilek, 2005). This verification is important because tumor deposited dose is dependent on source position accuracy. Thus, Monte Carlo simulations are a valuable approach for brachytherapy treatment planning systems. In this work a high dose rate Nucletron Microselectron V2 $^{192}$Ir radioactive source was simulated and data were validated by comparison with literature values. Simulations under brachytherapy treatment conditions with irradiation from a Nucletron Microselectron V2 $^{192}$Ir for measuring radioactive source position and dose were done in a 10 cm x 10 cm x 10 cm PMMA phantom. The absorbed dose is defined as the ratio between $d\bar{\epsilon}/dm$, where $d\bar{\epsilon}$ represents the mean energy imparted to a medium of mass $dm$, with unit of joule per kilogram (J/Kg) that is equivalent to gray (Gy) (Faiz, 1994; Podgorsak, 2003).

## Materials and Methods

Simulations have been performed with Monte Carlo N-Particle eXtended (MCNPX) version 2.3.0 developed at Los Alamos laboratory and deposited dose was evaluated with tally F6 (MeV/g/particle) with an Intel Core Duo T5600 1.83 GHz processor. Nucletron Microselectron V2 $^{192}$Ir radioactive source was implemented according to the new design microselectron-HDR $^{192}$Ir source dimensions described by Daskalov and colleagues (Daskalov, Löffler, & Williamson, 1998). All source geometry dimensions were respected but the tip and end of the stainless steel capsule were designed with a circular shape.

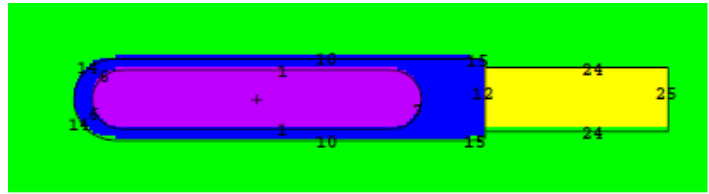

Figure 1 – $^{192}$Ir Nucletron Microselectron geometry implemented with MCNPX.

Stainless steel composition of capsule and cable is the one described in Table 1 with capsule density of 8.06 g cm$^{-3}$ and cable density of 4.81 g cm$^{-3}$. Radial dose and anisotropy functions were determined by placing $^{192}$Ir source in a water phantom with a radius of 15 cm in which the density of iridium and water are 22.4 g cm$^{-3}$ and 0.998 cm$^{-3}$, respectively. The photon energy spectrum of the $^{192}$Ir is shown in Figure 2 (based on energy values from (Cho, Muller-Runkel, & Hanson, 1999)).

Table 1- Stainless steel AISI 316L composition used for Monte Carlo simulations.

| Material | %C | %Mn | %Si | %Cr | %Ni | %P | %S | %Mo | %Fe |
|---|---|---|---|---|---|---|---|---|---|
| AISI 316L | 0.03 | 2 | 1 | 17 | 13 | 0.045 | 0.03 | 2.5 | 64.395 |

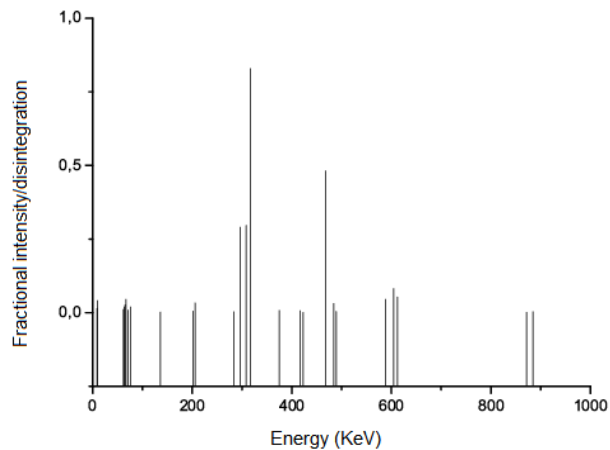

Figure 2 - Photon energy spectrum of $^{192}$Ir source.

## Results and discussion

Several dosimetry parameters can be determined allowing to compare results with those obtained by others. The absorbed dose rate can be expressed as:

$$\dot{D}(r,\theta) = S_k . \Lambda . \frac{G(r,\theta)}{G(r_0,\theta_0)} . g(r). F(r,\theta) \quad \text{Equation 1}$$

Where $S_k$ is the air kerma strength, $\Lambda$ is the dose rate constant, $G(r,\theta)$ is the geometry factor, $g(r)$ is the radial dose function, $F(r,\theta)$ is the anisotropy function at radial distance $r$ and angle $\theta$ being $(r_0,\theta_0)$ the reference point.

The radial dose function is expressed as:

$$g(r) = \frac{\dot{D}(r,\theta_0)G(r_0,\theta_0)}{\dot{D}(r_0,\theta_0)G(r,\theta_0)} \quad \text{Equation 2}$$

The anisotropy function is expressed as:

$$F(r,\theta) = \frac{\dot{D}(r,\theta)G(r,\theta_0)}{\dot{D}(r,\theta_0)G(r,\theta)} \quad \text{Equation 3}$$

Radial dose and anisotropy functions simulations enable to establish treatment planning systems in which calculated values can be used to evaluate dose distributions around [192]Ir radioactive sources. While radial dose function describes dose dependence with depth along the transverse axis of the source, the anisotropy function describes the anisotropy of dose distribution. A comparison of the obtained results for radial dose and anisotropy functions with those obtained by Daskalov et al. and Taylor et al. is done (Daskalov et al., 1998; Taylor & Rogers, 2008). Results obtained in this study are in agreement with those previously published. Radial dose function calculations were obtained for distances between 1 and 10 cm with $2 \times 10^9$ photons emitted by [192]Ir radioactive source (Figure 3). Absorbed dose detectors are cubic or cylindrical with a detection volume of 1 mm[3]. Anisotropy function was calculated with 10º angle intervals between 10º-170º and with 2º angle intervals between 0º-10º and 170º-180º (Figure 4). For anisotropy function calculation $5 \times 10^8$ photons were simulated and it was determined at $r = 0.5, 1, 2, 3, 5, 10$ cm. Dose detector has angular aperture of 2º and 0.1 cm of thickness and width. At angles near 0º and 180º the anisotropy function value is smaller (Figure 5). Factors that may influence possible differences between results obtained in this study and those obtained by others are phantom dimensions, different composition of simulated materials, distinct dimensions of dose detectors and different source code for Monte Carlo simulations. A better agreement is verified between results obtained in this study and those obtained by Daskalov and colleagues because simulations presented in this study were implemented based on [192]Ir source model described by Daskalov and colleagues.

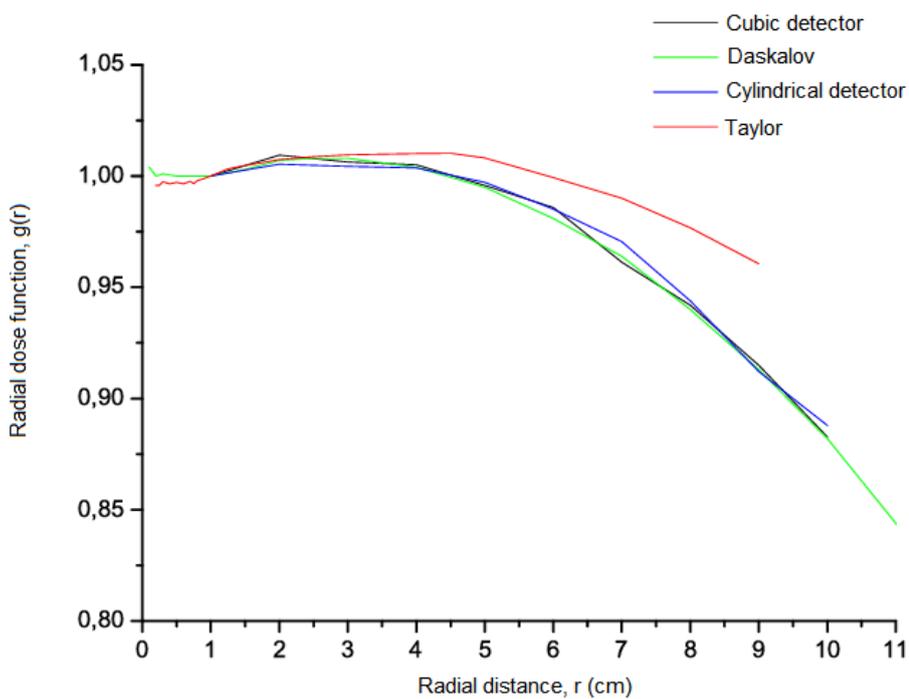

Figure 3 - Radial dose function. Results are shown for cubic and cylindrical detectors and compared with those from Daskalov et al. and Taylor et al. (Daskalov et al., 1998; Taylor & Rogers, 2008).

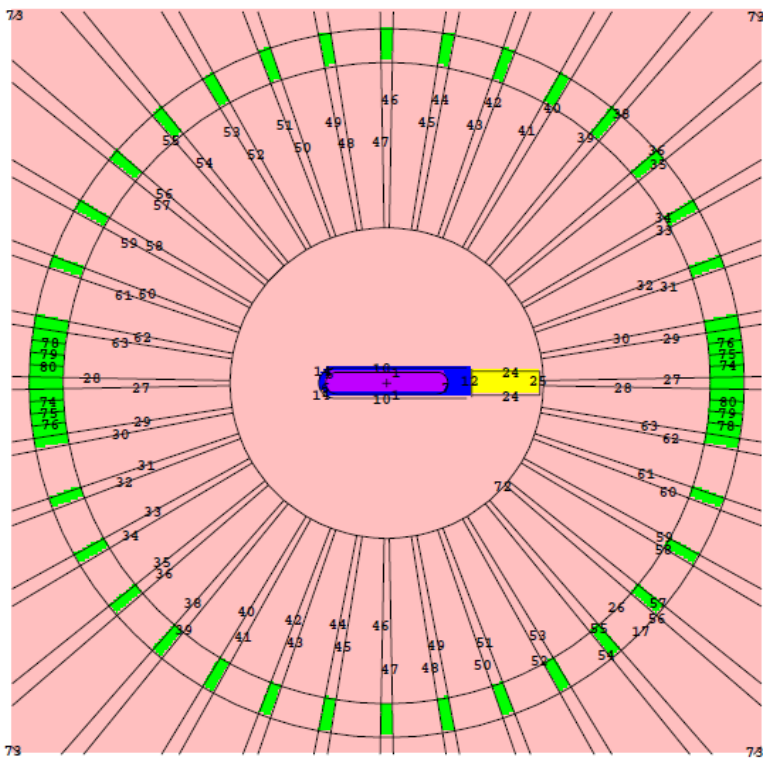

Figure 4 - Image of [192]Ir source geometry used for anisotropy function calculation at $r = 1\ cm$.

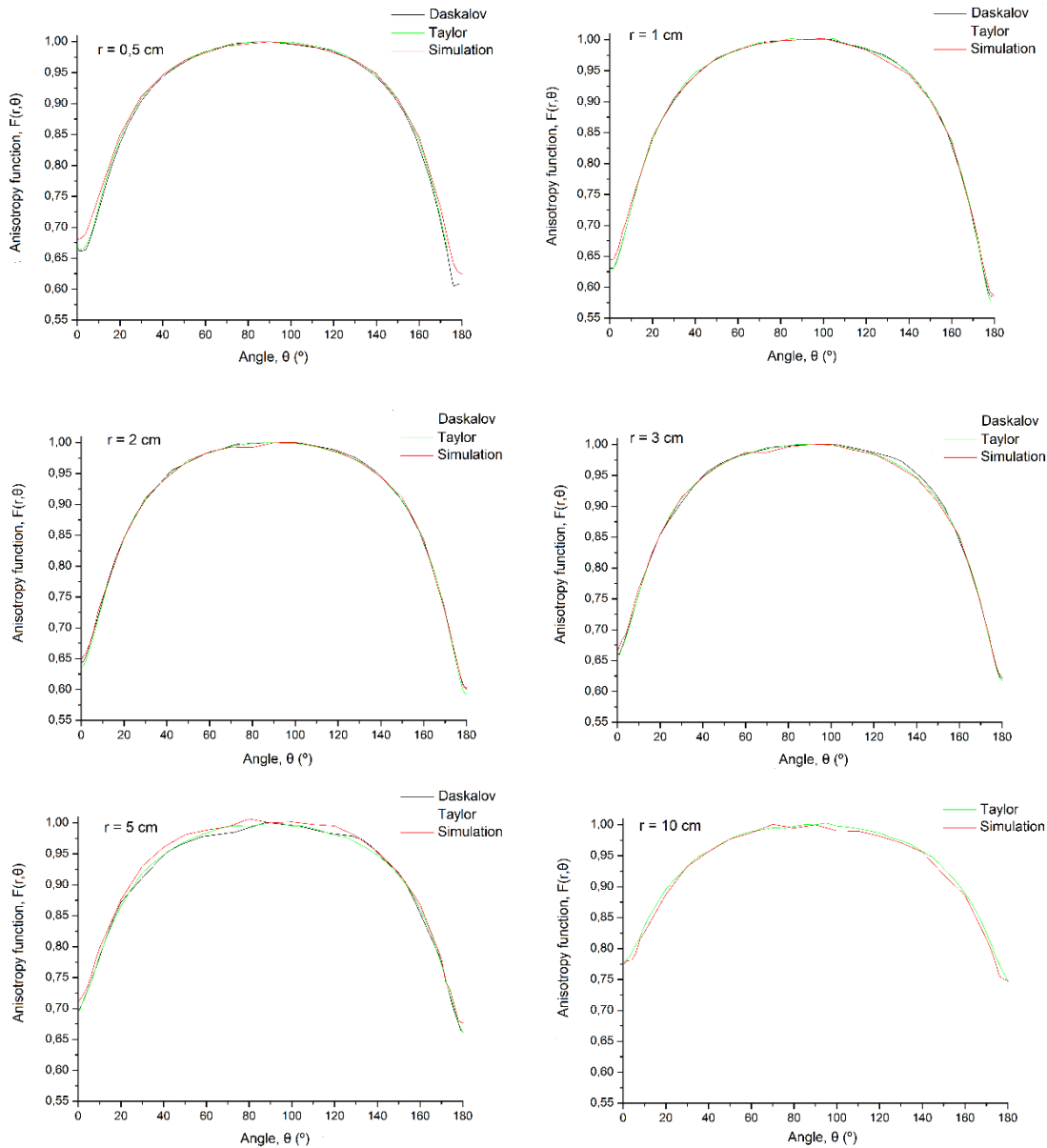

Figure 5 - Anisotropy function for several radial distances compared with results from Daskalov et al. and Taylor et al. (Daskalov et al., 1998; Taylor & Rogers, 2008).

Dose deposition usually follows the inverse square law (Podgorsak, 2003). At radial distances $r$ that are three times bigger than radioactive source length $L$ the source can be regarded as a point source. Thus, inverse square law can be used to characterize de dependence of absorbed dose with distance.

$$G(r,\theta)r^2 \approx 1 \; \forall_r > 3L \qquad \text{Equation 4}$$

Since radioactive source has active length of 3.6 mm and a dosimeter is placed at a distance of 1 cm the condition expressed in equation 4 is fulfilled (Sakelliou, Baltas, & Zamboglou, 2006). Therefore, radiation intensity detected with the dosimeter is determined by the inverse square law.

$$I \propto \frac{1}{r^2} \qquad \text{Equation 5}$$

According to Figure 6 $r^2 = d^2 + \left((n-n_0)\Delta x\right)^2$ where $r$ is radial distance, $d$ is the distance between the dosimeter and radioactive source in position $n$, $\left((n-n_0)\Delta x\right)$ represents source dwell position and $n_0$ represents the position in which dosimeter has a maximum response.

$$I \propto \frac{1}{d^2 + \left((n-n_0)\Delta x\right)^2} \qquad \text{Equation 6}$$

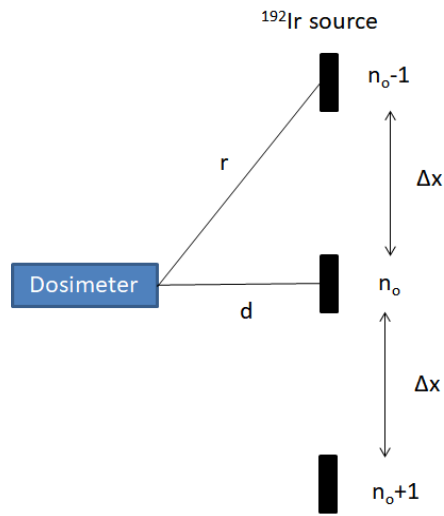

Figure 6 – Schematic representation of [192]Ir source position in relation to dosimeter, where $r$ is radial distance, $d$ is the distance between the dosimeter and radioactive source in position $n$.

After validating radial dose and anisotropy functions by comparison with data from other authors Monte Carlo simulations of [192]Ir source in a 10 cm x 10 cm x 10 cm PMMA phantom were performed. Brachytherapy source position moves in steps of 2 mm and is located inside a stainless steel needle as shown in Figure 7.

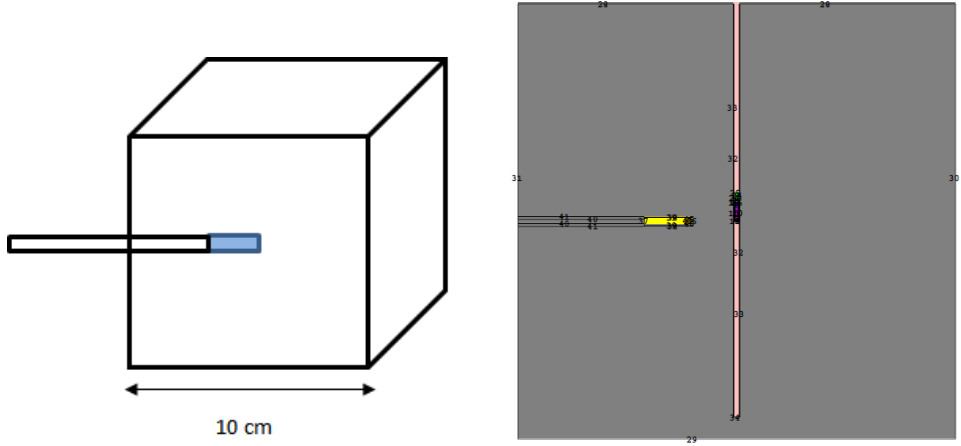

Figure 7 - Source position in a PMMA phantom in which [192]Ir source is located inside brachytherapy needle.

Brachytherapy needle has a length of 16 cm, inner wall thickness of 0.15 mm and external diameter of 1.5 mm. Simulations take into account inhomogeneities due to air inside the needle that has a density of 0.00120429 g cm[-3] and the composition described in Table *2*

Table 2 - Air composition used for Monte Carlo simulations.

| % C | % N | % O | % Ar |
|---|---|---|---|
| 0,0124 | 75,5267 | 23,1781 | 12,827 |

The simulated PMMA phantom has a density of 1.18 g cm[-3] and the dosimeter is a luminescent probe made of PMMA cladding with density of 1.2 g cm[-3] and polystyrene core with density of 1.05 g cm[-3]. The luminescent material is connected to a PMMA optical fiber that transports the luminescent signal to a photodetector enabling to determine absorbed dose. Simulation was done with $1 \times 10^8$ photons and by using equation 6 with $d = 1\ cm$ and $\Delta x = 0.2\ cm$ a theoretical fit has been obtained to Monte Carlo calculations of tally F6 as can be seen in Figure 8.

Dose values can be estimated knowing the activity of [192]Ir radioactive source. This can be done by converting dose calculation obtained with tally F6 to dose rate values:

$$\dot{D}(r) = Dose \frac{\left(\frac{MeV}{g}\right)}{particle} * 1.602E - 10 \left(\frac{J.g}{Mev.Kg}\right) * A \left(\frac{particle}{s}\right) \qquad \text{Equation 7}$$

Where $\dot{D}(r)$ represents dose rate at radial distance $r$ and $A$ is radioactive source activity. Dose rate calculations obtained with MCNP are shown in Table 3.

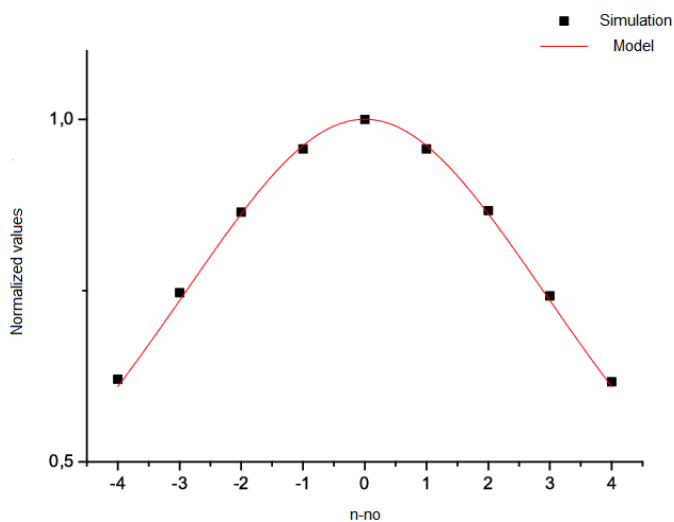

Figure 8 - Theoretical fit to Monte Carlo calculations of tally F6 for various $^{192}$Ir source positions.

Table 3 – Dose rate calculations for dosimeter with two distinct lengths.

| Activity (Ci) | $\emptyset$ $(mm)$ | $L$ $(cm)$ | $\dot{D}$ $(\frac{cGy}{s})$ |
|---|---|---|---|
| 8.9 | 2 | 1 | 3.58 |
| 6.81 | 2 | 0.5 | 2.75 |

High dose rate brachytherapy treatments deliver a dose rate superior to 12 Gy/h that is equivalent to 3.33 cGy/s (Podgorsak, 2003). Therefore, dose rate values obtained are acceptable. As source activity decays as time progresses dose rate also decays.

**Conclusions**

Monte Carlo computational simulation of $^{192}$Ir radioactive source was performed with MCNPX version 2.3.0. After validating radial dose and anisotropy functions by comparison with data from Daskalov et al. and Taylor et al. results show that it is possible to determine $^{192}$Ir source position by using equation 6 and enabling to plan treatments according to patient requirements (Daskalov et al., 1998; Taylor & Rogers, 2008). Moreover, these simulations enable to establish a fiber optic dosimeter as a tool for brachytherapy quality control. This device is capable of determining absorbed dose and position of the brachytherapy source, as assessed by Monte Carlo simulations.